\documentclass[twocolumn,twocolappendix]{aastex631}

\newcommand\bmm{{$\beta$-meteoroid }}
\newcommand\bms{{$\beta$-meteoroids }}
\newcommand\amsn{{$\alpha$-meteoroids}}
\newcommand\amm{{$\alpha$-meteoroid }}
\newcommand\ams{{$\alpha$-meteoroids }}
\newcommand\bmsn{{$\beta$-meteoroids}}
\newcommand\bs{{$\beta$-stream }}
\newcommand\bsn{{$\beta$-stream}}

\newcommand\rs{R_\odot}
\newcommand\rst{$\rs$}

\usepackage{mathtools}
\usepackage{amsmath}
\usepackage{amssymb}
\usepackage{rotating}
\usepackage{latexsym}

\usepackage[finalnew]{trackchanges}
\submitjournal{PSJ}

\shorttitle{Size distribution of small zodiacal grains}
\graphicspath{{./}{}}

\begin{document}

\title{Size distribution of small grains in the inner zodiacal cloud}

\correspondingauthor{Jamey Szalay}
\email{jszalay@princeton.edu}

\author[0000-0003-2685-9801]{J. R. Szalay}
\affil{Department of Astrophysical Sciences, Princeton University, 171 Broadmead St., Princeton, NJ 08540, USA}

\author[0000-0002-5667-9337]{P. Pokorn\'y}
\affiliation{Astrophysics Science Divison, NASA Goddard Spaceflight Center, Greenbelt, MD, 20771, USA}
\affiliation{Department of Physics, The Catholic University of America, Washington, DC, 20064, USA}

\author[0000-0003-1191-1558]{D. M. Malaspina}
\affiliation{Department of Astrophysical and Planetary Sciences, University of Colorado Boulder, Boulder, CO, USA}
\affiliation{Laboratory for Atmospheric and Space Physics, University of Colorado Boulder, Boulder, CO, USA}



\begin{abstract}
The Parker Solar Probe (PSP) spacecraft has transited the inner-most regions of the zodiacal cloud and detects impacts to the spacecraft body via its electric field instrument. Multiple dust populations have been proposed to explain the PSP dust impact rates. PSP's unique orbit allows us to identify a region where the impact rates are likely dominated by \amsn, small zodiacal grains on approximately circular, bound orbits. From the distribution of voltage signals generated by dust impacts to PSP in this region, we find the cumulative mass index for grains with radii of $\sim$0.6-1.4 $\mu$m (masses of $3\times10^{-15}$ to $3\times10^{-14}$ kg) to be \change{$\alpha = 1.3 \pm 0.2$}{$\alpha = 1.1 \pm 0.3$} from 0.1-0.25 \rst. $\alpha$ increases \remove{monotonically} toward the Sun, with even smaller \remove{collisional} fragments generated closer to the Sun. The derived size distribution is steeper than previously estimated, and in contrast to expectations we find most of the dust mass resides in the smallest fragments and not in large grains \add{inside 0.15 au}. As the inner-most regions of the zodiacal cloud are \add{likely} collisionally evolved, these results place new constraints how the solar system's zodiacal cloud and by extension astrophysical debris disks are partitioned in mass.

\end{abstract}

\keywords{}


\section{Introduction} \label{sec:intro}

Our solar system's zodiacal cloud comprises multiple dust populations. Of the interplanetary populations, these are typically categorized as \amsn\footnote{There are two definitions of \ams in the literature \citep{sommer:23a}: 1) a dynamical subset of gravitationally bound grains with very large eccentricities and 2) all grains gravitationally bound to the Sun. We use the latter, broad definition for \amsn.}: gravitationally bound grains on elliptic trajectories and \bmsn: grains so small the outward force of solar radiation overcomes the inward pull of the Sun's gravity such that they are pushed out of the solar system on hyperbolic trajectories \citep{zook:75a, wehry:99a}. These dust populations are continuously evolving and exchanging material due to dynamical evolution from gravity, solar radiation, and electromagnetic forces, as well as disruption from collisions, sublimation, and rotation \citep{mann:04a}. This cloud also provides an important analog to study exozodiacal systems. 



A key quantity that directly affects \bmm production and zodiacal erosion is how collisional fragment particles are distributed in mass, which has not been well constrained near the Sun. Collisional products are often assumed to follow a power-law mass distribution \citep[e.g.][]{gault:69a,dohnanyi:69a}, \add{such that the immediate fragmented differential mass distribution is $f(m) \propto m^{-(\eta+1)}$. These collisional products can then undergo mass-dependent transport and erosion, such that the instantaneous differential mass distribution at any location in the heliosphere has the form}
\begin{equation}
f(m) \propto m^{-(\alpha+1)}
\end{equation}
where $\alpha$ is the cumulative mass index. \add{In this study, we assume the cumulative fragmentation index $\eta$ and the cumulative mass index $\alpha$ are equal, such $\eta = \alpha$.} Note, $\alpha$ as the variable used to represent mass index does not have any relation to the term `$\alpha$-meteoroid', where we use `$\alpha$' in both cases for historical consistency. Changes in $\alpha$ directly affect how much material the zodiacal cloud retains or sheds through its production of \bmsn. Hence, an accurate determination of $\alpha$ reveals the effect of collisional fragmentation on zodiacal evolution.  

There have been a number of both observational and ground-based experimental constraints on $\alpha$. Ground-based collision experiments found values of $\alpha = 0.5-1.0$ for collisionally generated material \citep[e.g.][and refs. therein]{krivov:00a}. The size distribution of lunar ejecta in the radius range of $0.3-10$ $\mu$m was consistent with $\alpha = 0.9$ \citep{horanyi:15a}, while initial Solar Orbiter (SolO) results \citep{zaslavsky:21a} found values of $\alpha = 0.3-0.4$ for \bms with radii of $\sim$100-200$~\mu$m.  The zodiacal cloud in the outer solar system beyond Saturn's orbit as measured by New Horizons' Student Dust Counter observations \citep{piquette:19a} was found to be \add{consistent with} $\alpha = 0.5-1.2$ \add{by comparing with a dynamical model of the outer zodiacal grain distribution} \citep{poppe:16c}. More recent collisional modeling in the inner solar system concluded $\alpha = 1.11 \pm 0.03$ for grains with radii $>100$ $\mu$m \citep{pokorny:24a}. From modeling the near-Sun zodiacal cloud, $\alpha = 0.5-0.6$ was expected at a heliocentric distance of 0.1 au for micron-sized grains \citep{ishimoto:98a}. 


While unequipped with a dedicated dust detector, Parker Solar Probe (PSP) spacecraft \citep{fox:16a} registers dust impacts primarily via voltage measurements with the FIELDS instrument \citep{bale:16a}.  There has been an extensive history of measuring dust impacts from spacecraft equipped with electric field instruments, for example: Voyager 2 \citep{gurnett:83a}, Vega \citep{laakso:89a}, DS-1 \citep{tsurutani:03a,tsurutani:04a}, Wind \citep{malaspina:14a,kellogg:16a,malaspina:16a}, MAVEN \citep{andersson:15a}, \add{Cassini} \citep{ye:2014a}, STEREO \citep{zaslavsky:12a,malaspina:15a} and MMS \citep{vaverka:18a,vaverka:19a}.


During its first three orbits, impact rates observed by PSP were consistent with fluxes of dominantly high-speed, submicron-sized \bms leaving the solar system on escaping orbits \citep{szalay:20a,page:20a,malaspina:20a}. \add{Grains smaller than} \bmsn, \add{often termed nanograins, are less susceptible to radiation pressure and their dynamics are dominated by electromagnetic forces; these} nanograins \remove{which experience strong electromagnetic forces} were not found to be dominantly responsible for the impact rates during PSP's 2nd orbit \citep{mann:20a}.  A two-component model consisting of both \ams and \bms was subsequently used to explain the first six orbits of impact rate data \citep{szalay:21a}, which was able to well-reproduce the majority of the impact rates observed. Additionally, a PSP dust database has been published which is continually updated as more data is taken throughout the PSP mission \citep{malaspina:23a}. 


In this study, we will calculate the cumulative impact amplitude distributions for all PSP-derived dust impact rates \citep{malaspina:23a}. Comparing to the existing two-component model, we identify a region in a subset of PSP orbits dominated by \ams in the heliocentric distance range of $0.1-0.25$ au and fit the impact amplitude distributions to a power-law. This will allow us to quantitatively constrain the size distribution of small, bound \ams in the near-Sun environment.



\section{PSP dust impact rates from orbits 1-17}

\begin{figure*}
\plotone{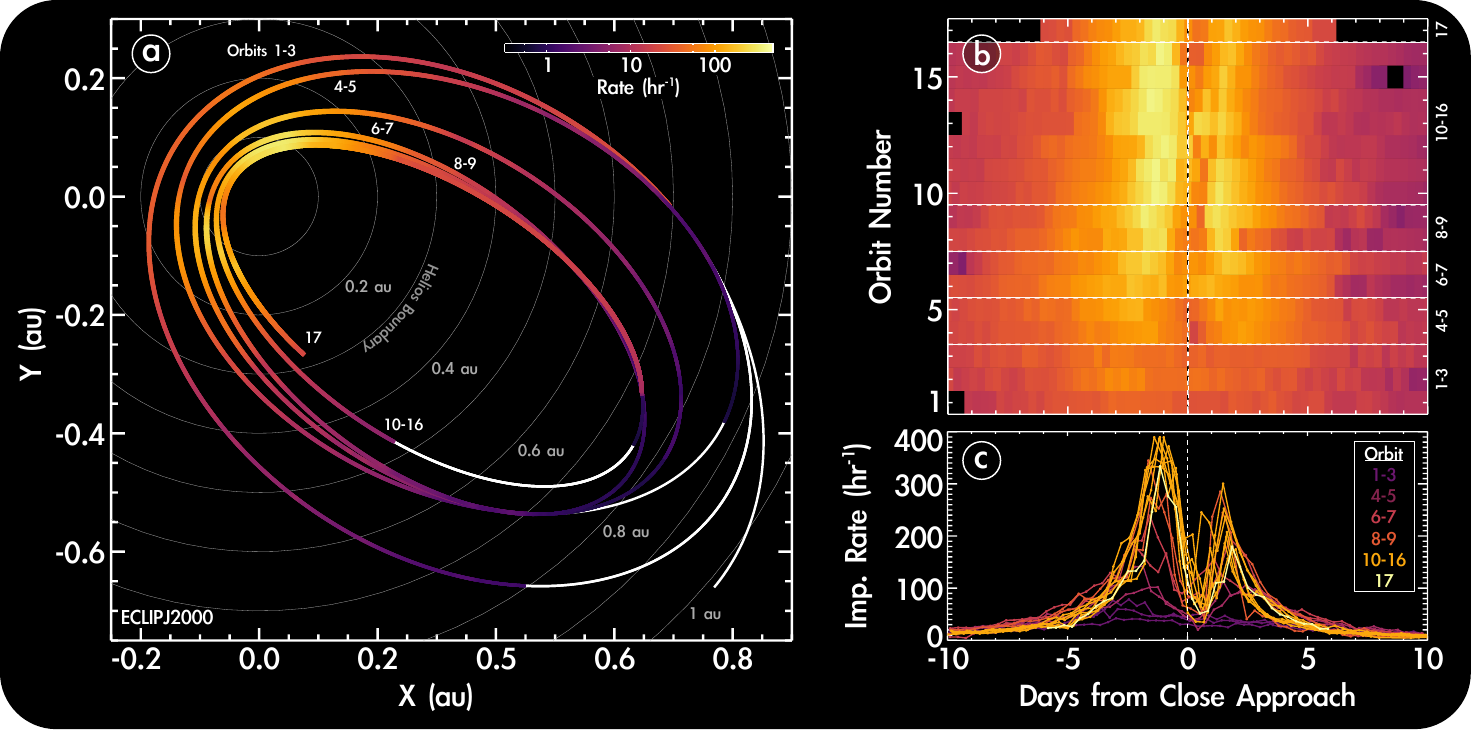}
\caption{Impact rates for PSP's first 17 orbits. (a) Top-down view in Ecliptic J2000 coordinates, where rates within each orbital group are averaged. The inner-most location of the Helios dust measurements \citep{grun:80a} outside 0.3 au is also indicated. (b, c) Time-series of each orbit, where the colorbar in (b) is shared with that of (a), as a function days from perihelion. \label{fig:rate}}
\end{figure*}

\begin{figure*}
\plotone{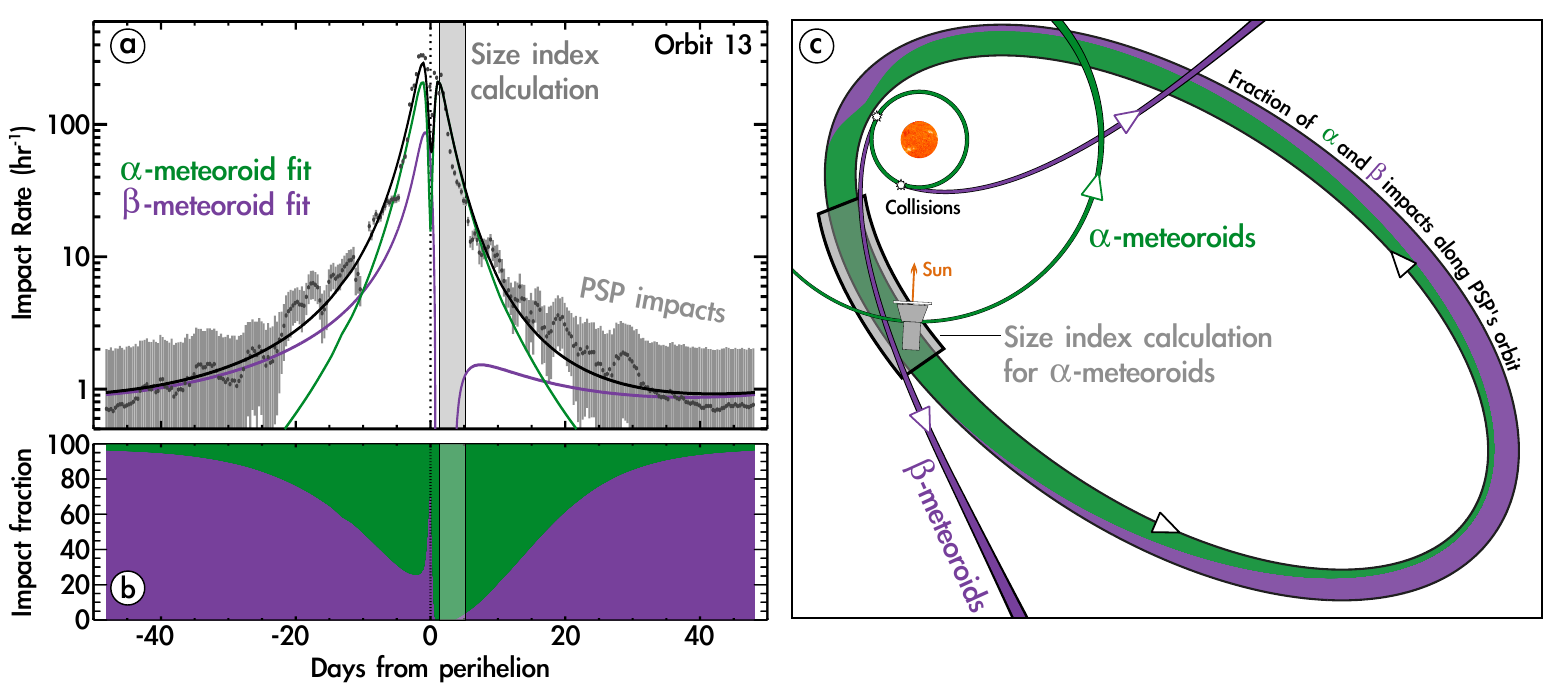}
\caption{(a) Example impact rate from orbit 13, representative of orbits 10-16, along with the modeled abundances of \ams and \bms from our two-component fit \citep{szalay:21a}. (b) Fractional contribution of \ams and \bms to total modeled impact rate. (c) Same information as panel (b) shown in the ecliptic J2000 x-y plane with example \amm \& \bmm trajectories. \label{fig:ratio}}
\end{figure*}

Figure~\ref{fig:rate} shows the dust impact rates \citep{malaspina:23a} to PSP as a function of position and time from orbits 1-17. There are six orbital groupings for the spacecraft shown: 1-3, 4-5, 6-7, 8-9, 10-16, \& 17. There are a few notable features in the impact rates: all orbits exhibit a peak pre-perihelion, with lower impact rates observed at perihelion, followed often by a post-perihelion peak that is lower in total rate than the pre-perihelion peak. A two-component model consisting of \change{pro-grade}{prograde} \ams and \bms was used to successfully model PSP impact rates from orbits 1-6 \citep{szalay:21a}. Note, this model incorporates the updated density profile of the zodiacal cloud derived from PSP imaging \citep{stenborg:20a}. Figure~\ref{fig:ratio}a shows an example of this two-component model fit to the impact rates from orbit 13 as a representative example. The model predicts a symmetric pre- and post-perihelion impact rate peak from the \amm population (green line) and an asymmetric profile for the \bmm population. Due to the spacecraft velocity addition with the solely outbound, prograde \bmsn, the impact rate from \bms is expected to exhibit a strong pre-perihelion peak and significantly depleted impact rates just after perihelion due to PSP ``catching up'' to the \bms and having \change{very small}{much smaller} relative impact speeds and corresponding impact rates. 

The two-component model successfully fits the majority of impact rate structure observed by PSP, with the exception of the post-perihelion peak magnitude for orbits 4-5. During these orbits, which exhibited a post-perihelion impact rate peak that could not be explained by the two-component model, it was proposed that a third component of collisionally produced \bms from the Geminids stream, termed a \bsn, could be responsible for the additional enhancement \citep{szalay:21a}. It was shown that PSP would have intersected the trajectories of \bms produced during collisions between the Geminids and zodiacal cloud in orbits 4-5.  Subsequent directional analysis was also supportive of a \bs hypothesis \citep{pusack:21a}. Modeling efforts on the overall Geminids stream found the bound \ams within the Geminids could not be responsible for the additional rate enhancement \citep{cukier:23a}, indirectly supporting the \bs hypothesis. However, without dedicated dust instrumentation it is difficult to definitively conclude the origins of the post-perihelion enhancement during orbits 4-5.

Aside from orbits 4-5, the model is able to well-reproduce the magnitude and location of the post-perihelion enhancement. This region, just after perihelion, is the primary focus of this study as it is expected to be dominated by impacts from \amsn, with minimal contributions from \bmsn. Figures~\ref{fig:ratio}b \& c show the impact fraction of \ams (green) and \bms (purple) throughout orbit 13. As shown here, the region just after perihelion to $\sim$5 days post-perihelion has less than a few percent contribution from \bmsn. Therefore, due to PSP's highly eccentric orbit, this region provides a unique opportunity to isolate the \amm population and analyze its properties. We focus on this region throughout the remainder of this study and derive the size distribution power-law index for \ams here. Additionally, we focus on orbits 10-16 as they provide seven nearly identical trajectories to sample this region and provide a statistical dataset with which to study \ams in the near-Sun zodiacal cloud.

\section{Determining the Size Distribution} \label{sec:dist}

\add{Here, we outline a method to derive the cumulative mass index $\alpha$ from the distribution of impact voltages to PSP. } \add{Dust impacts are measured by PSP as a potential change $\Delta V \propto Q_\mathrm{imp}$} \citep{collette:14b,shen:23a}. \add{In order to relate the impact charge distribution to the impact mass distribution, we assume the differential charge distribution has the form $f(Q) \propto Q^{-(\alpha_Q+1)}$, where $\alpha_Q$ is the cumulative charge index. In the Appendix, we derive a relation between $\alpha_Q$ and $\alpha$ that allows us to estimate the mass index from the distribution of measured impact voltages as similarly done for radar observations of meteors with a speed-dependent radar amplitude} \citep{pokorny:16a}.




\begin{figure*}
\plotone{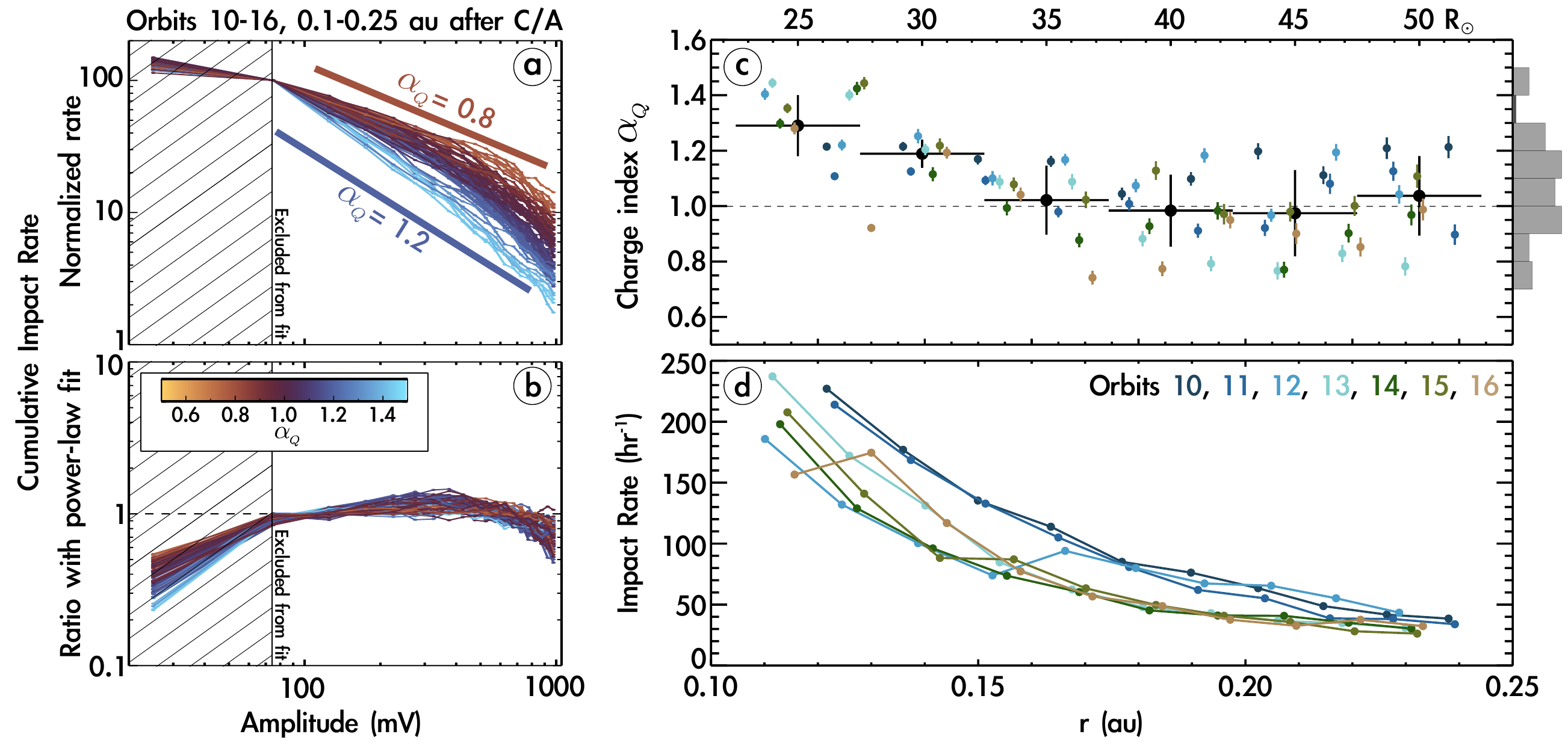}

\caption{ (a) Cumulative impact rate distribution as a function of impact amplitude for orbits 10-16, with color indicating the power-law fit index above 50 mV for each 8-hr impact rate interval considered. b) Ratio of cumulative impact rates with their corresponding power-law fits.(c) Cumulative charge indices $\alpha_Q$ as a function of heliocentric distance, with average values and standard deviation each 5 $\rs$ and the histogram on the right shows the distribution of $\alpha_Q$ values.   (d) Impact rates as a function of heliocentric distance. Colors for (c) \& (d) show orbit number \label{fig:index}}
\end{figure*}


To ensure PSP is being impacted by a single population, we focus on times where the spacecraft resides between 0.1 to 0.25 au (22-54 $\rs$) post-perihelion, as the two-component model predicts \ams contribute $>$$95 \%$ of the total impacts throughout this region for orbits 10-16. \remove{Figure~3a shows the cumulative impact amplitude distribution measured by PSP in this region, normalized to the impact rates at 100 mV amplitude.}. We note interstellar grains could also be impacting PSP in this region \citep{sterken:12a, strub:19a}, where due to PSP's orbit geometry these grains would be most readily detected during the post-perihelion arc far from the Sun. However, given the ability of the two-component model to reasonably reproduce the location and amplitude of the second peak during orbits 10-16 and that much of the interstellar dust grains are dynamically prevented from transiting very near to the Sun due to radiation pressure, we do not expect this source to appreciably contribute to PSP's impact rates in this region.

Figure~\ref{fig:index}a \change{shows the distribution of cumulative impact rates investigated.}{shows the ensemble of cumulative impact-amplitude distributions measured by PSP for each 8-hour segment within the selected portions of orbits 10-16. Each voltage range is displayed at the center of a 50 mV bin, such that the first voltage point in the shaded region corresponds to 0-50 mV, the second point to 50-100 mV, and so forth.} Due to the dust impact identification method, impacts with amplitudes below \change{100}{50} mV \add{(the first point in each curve)} are more challenging to detect \citep{malaspina:23a} and we are likely not identifying all impacts below this amplitude, hence the break in the slope of the distribution at this amplitude as shown in Figure~\ref{fig:index}a. Additionally, this measurement technique saturates above \change{100}{950} mV \remove{(1 V)}, therefore we can probe the size-distribution index within a single decade in mass.  We fit the cumulative amplitude distribution above \change{100}{50} mV to a power-law, where the colors of the various curves in Figure~\ref{fig:index}a shows the cumulative charge index \change{$\alpha$}{$\alpha_Q$}. Figure~\ref{fig:index}b \add{shows these same cumulative impact rate curves divided by their corresponding power-law fits.} As shown here, the cumulative voltage distributions are highly consistent with power-laws. In Figure~\ref{fig:index}c, we show the fitted \change{mass}{charge} indices as a function of heliocentric distance, along with average values in 5 $\rs$ bins. Figure~\ref{fig:index}d shows the total impact rates for each orbit, where Figures~\ref{fig:index}c \& d are colored by orbit number.

We previously estimated the minimum detectable bound \amm grain radius to be approximately 0.6 $\mu$m for a threshold of \change{100}{50} mV \citep{szalay:21a}. \add{With the additional uncertainty discussed in the Appendix, this lower limit could be down to $\sim$0.5 $\mu$m, however, grains as small as 0.5 $\mu$m in radius are expected to have $\beta \approx 0.5$ and would be unbound. Therefore, we retain our lower limit estimate to be 0.6 $\mu$m} As the PSP impact voltages used here have a dynamic range of \change{$100 - 1000$}{$50-950$} mV, we are able to assess the mass distribution within a decade in mass from the minimum detectable size such that the derived indices are relevant for grains with radii of approximately $0.6 - 1.4$ $\mu$m.

\begin{figure*}
\plotone{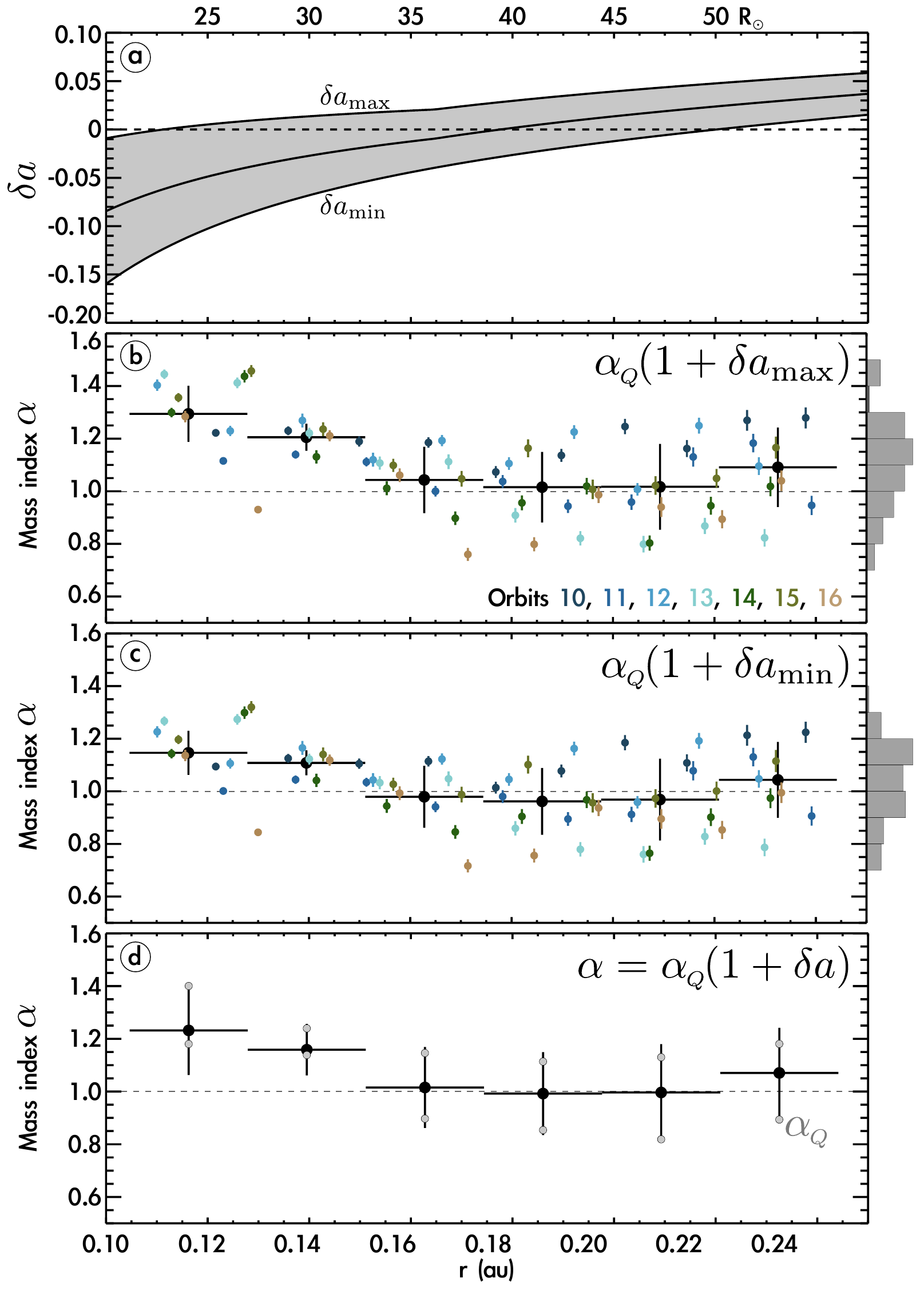}
\caption{ (a) Correction factor $\delta a$ for the mass-dependent portion of the impact speed. (b) \& (c) Corrected cumulative mass indices $\alpha = \alpha_Q(1+\delta a)$ corresponding to the minimum and maximum $\delta a$ values from (a). (d) Comprehensive estimate of $\alpha$ accounting for mass-dependent biases, with the uncorrected $\alpha_Q$ ranges shown with grey dots. \label{fig:alpha}}
\end{figure*}

\add{As derived in the Appendix, $\alpha = \alpha_Q(1+\delta a)$ relates the cumulative charge index $\alpha_Q$ to the cumulative mass index $\alpha$.} Figure~\ref{fig:alpha}a \add{shows $\delta a$ as a function of heliocentric distance, where this term is most enhanced for closer heliocentric distances. We use these $\delta a$ ranges to correct the derived cumulative charge indices $\alpha_Q$ from} Figure~\ref{fig:index}c. \add{Considering both the upper and lower limits of $\delta a$,} Figures~\ref{fig:alpha}b \& \ref{fig:alpha}c \add{show the range of corrected mass indices. For each $5R_\odot$ heliocentric bin, we determine the full range of possible $\alpha$ values from these two panels and display it in panel (d) with the average value of these ranges shown with black circles. The grey circles overlaid on these show the uncorrected ranges of $\alpha_Q$ from} Figure~\ref{fig:index}c. \add{While this mass-dependent correction is important to consider, we find it does not have a substantial effect on the dependence of $\alpha$ on $\alpha_Q$.}

\add{The mass indices shown in} Figure~\ref{fig:index}d, \add{represent our best estimate of $\alpha$ correcting for  mass-dependent biases discussed. Throughout this region, from the full uncertainty ranges in} Figure~\ref{fig:index}d, we find \change{$\alpha = 1.3 \pm 0.2$}{$\alpha = 1.1 \pm 0.3$}. Additionally, we find the mass index is higher near the Sun, exhibiting an enhancement in $\alpha$ inside $\sim$0.15 au systematically above \change{the average}{1.0}.


\section{Implications for collisional production} \label{sec:implications}

\begin{figure}
\plotone{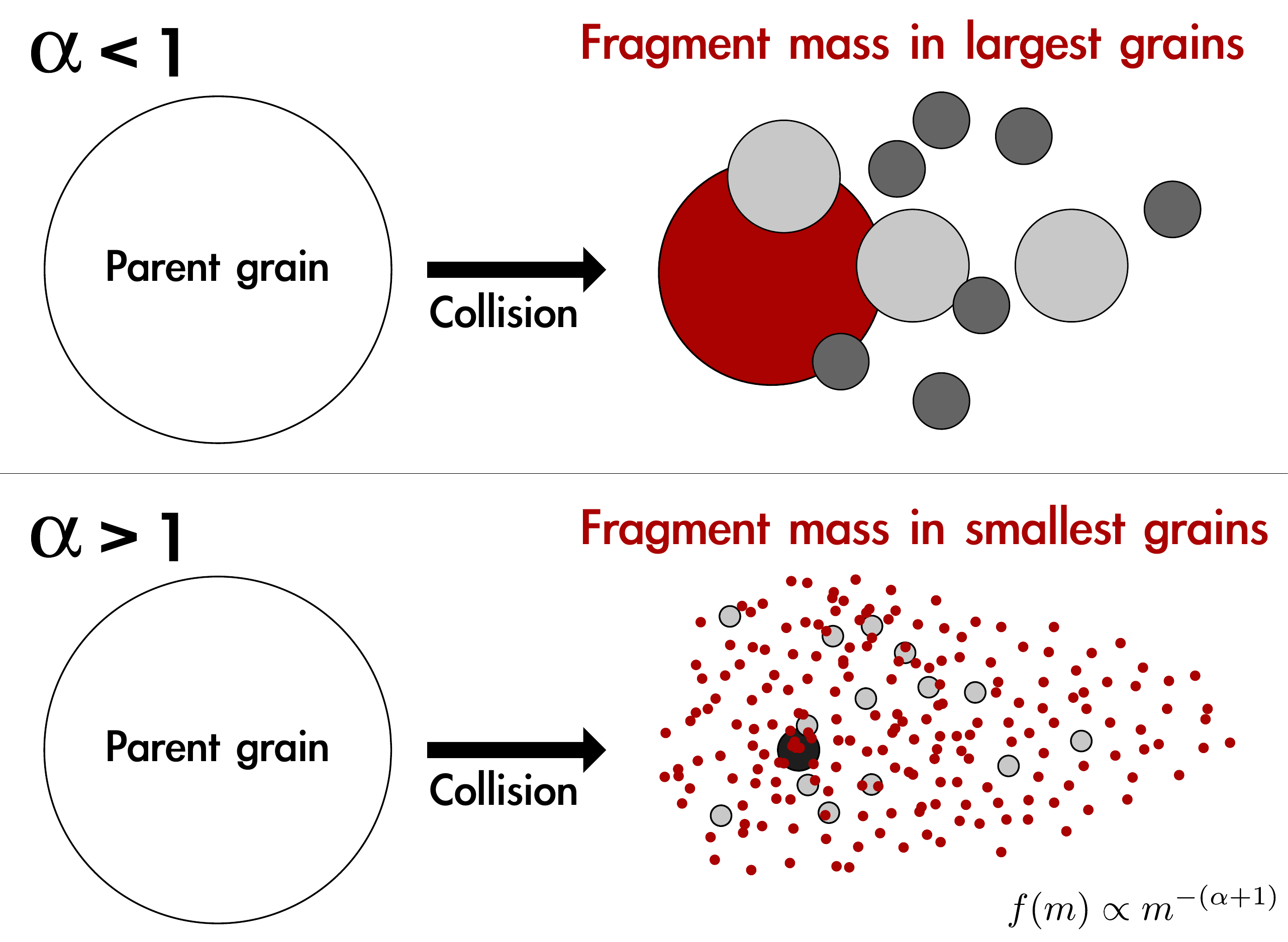}
\caption{Distribution of fragmentation products depending on the cumulative mass index. Fragments with $\alpha > 1$ have most of the mast distributed into small grains, while for $\alpha < 1$ the majority of mass is distributed into the few largest fragments. \label{fig:collision}}
\end{figure}

Having found \change{that on average $\alpha = 1.3 \pm 0.2$}{$\alpha = 1.1 \pm 0.3$} in the 0.1-0.25 au range, we now focus on the implications of these mass indices in the near-Sun zodiacal dust cloud. For a collisionally produced population, the mass index \remove{has physical intuitive meaning as it} determines the maximum size grain that can be produced as a fragment. We investigate how mass is distributed, including the maximum fragment size, as a function of $\alpha$. For an upper size cutoff $m_c$ for fragments, we normalize the distribution such that the total number of particles with mass greater than \add{or equal to} $m_c$ is 1. Therefore, the differential distribution function is
\begin{equation}
f(m) =     \frac{\alpha}{m_c^{-\alpha}-m_1^{-\alpha}} m^{-(\alpha+1)}  \label{eq:mdiff}
\end{equation}
where the number of particles between $m_a$ and $m_b$ is given by
\begin{equation}
N =  \int_{m_a}^{m_b} f(m) \ dm = \frac{m_a^{-\alpha}-m_b^{-\alpha}}{m_c^{-\alpha}-m_1^{-\alpha}} 
   \label{eq:n}
\end{equation}
For a parent particle of mass $m_1$, whose minimum fragment  mass is $m_0$, its collisional fragments must have a total mass of $m_1$, hence,
\begin{equation}
m_1 = \int_{m_0}^{m_1} m f(m) \ dm = \frac{\alpha}{m_c^{-\alpha}-m_1^{-\alpha}} \int_{m_0}^{m_1} m^{-\alpha} \ dm
\end{equation}
Therefore, the mass of the largest fragment $m_c$ is given by
%
%
\begin{equation}
m_c = 
    \begin{cases}
     \left( \frac{(\alpha-1)m_1 }{\alpha m_0^{1-\alpha}-m_1^{1-\alpha}}  \right)^{\tfrac{1}{\alpha}} & \text{if } \alpha \ne 1,\\
     \frac{m_1}{1+\ln{\left( m_1/m_0 \right)}} & \text{if } \alpha = 1
    \end{cases}
    \label{eq:mc}
\end{equation}
For a more intuitive interpretation, these relations can be approximated by assuming $m_0 \ll m_1$ and expanding to zeroth order in $m_0/m_1$ when $\alpha \ne 1$,
%

\begin{equation}
m_c \approx
    \begin{cases}
     m_0 \left( \frac{\alpha-1}{\alpha} \frac{m_1}{m_0} \right)^{\tfrac{1}{\alpha}} & \text{if } \alpha > 1,\\
     m_1 (1-\alpha)^{\tfrac{1}{\alpha}}  & \text{if } \alpha < 1
    \end{cases}
    \label{eq:mc_approx}
\end{equation}
As shown in Equations \ref{eq:mc} and \ref{eq:mc_approx}, for $\alpha > 1$ the maximum fragment cutoff mass is most strongly governed by $m_0$ and weakly depends on the ratio $m_1/m_0$, while for $\alpha \leq 1$ this cutoff mass is entirely governed by the original grain size $m_1$. For $\alpha = 1$ it is an intermediate case mostly governed by $m_1$ and weakly depending on $m_1/m_0$. This also means for steeper distributions with $\alpha > 1$, most of the material is in small grains near the lower end of the mass range, while shallower distributions with $\alpha \leq 1$ have most of the material in large grains toward the upper end of the mass range. Figure~\ref{fig:collision} \add{shows a visualization of these two regimes}.

\begin{figure}
\plotone{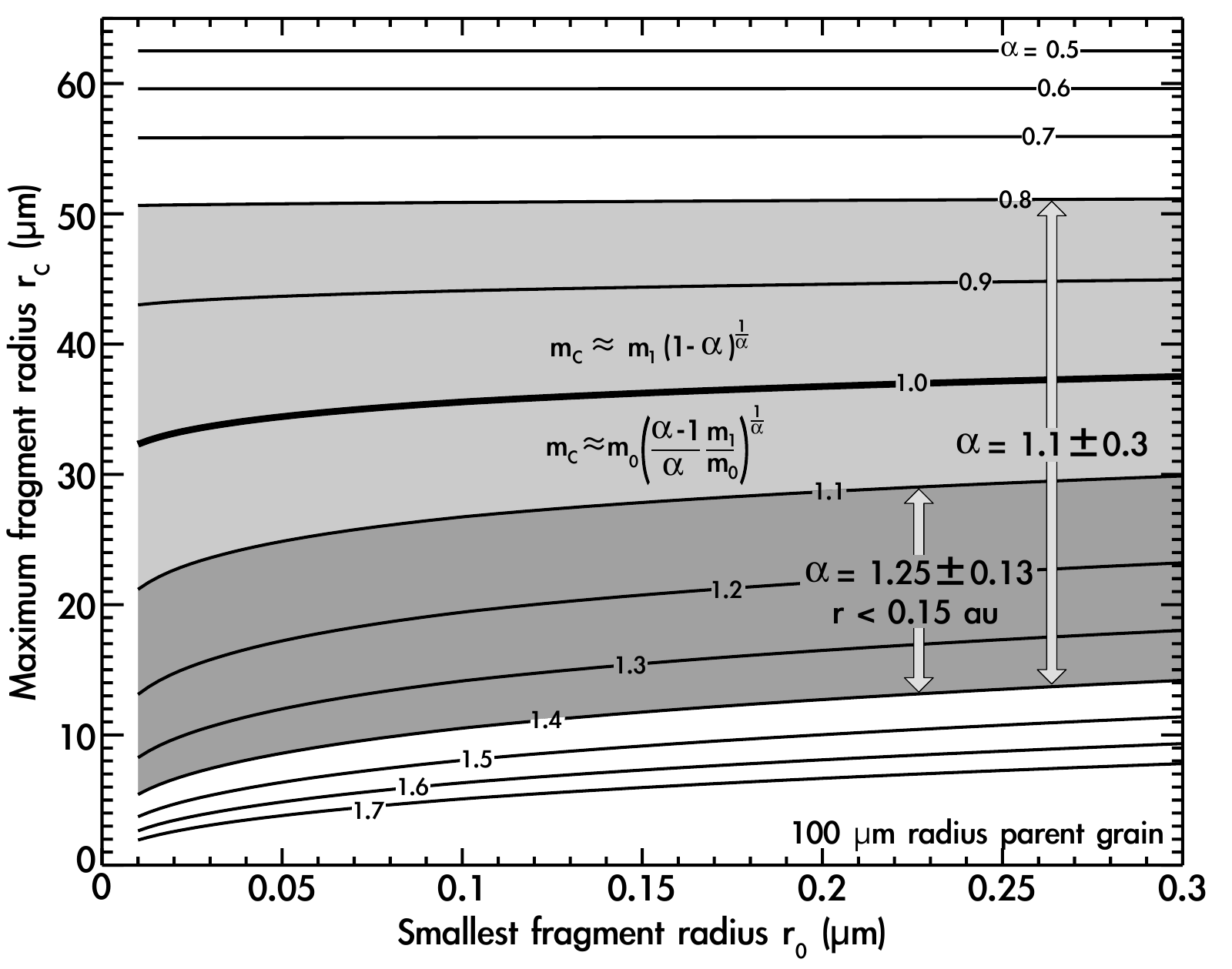}
\caption{Maximum fragment cutoff size as a function of minimum fragment size and cumulative mass index $\alpha$. The shaded region shows the average cumulative mass indices found to be consistent with the near-Sun environment. \label{fig:mc}}
\end{figure}

Figure~\ref{fig:mc} provides a visualization of these relations. It shows the maximum fragment size $r_c$ corresponding to grains with mass $m_c$ as a function of assumed smallest fragment size $r_0$ from 10 nm to 0.3 $\mu$m corresponding to mass $m_0$ for different values of $\alpha$. To determine radius for a given mass, we assume grains are ``old cometary'' origin according to \citet{wilck:96a}, \add{with grain mass densities in the range of 2.0-3.3 g cc$^{-1}$ for grain radii of 10 nm to 20 $\mu$m}. For $\alpha < 1.0$, the dependence of maximum fragment size on $r_0$ is weak, as most of the mass is partitioned amongst the largest grains. For $\alpha > 1.0$, most of the mass is partitioned to the smallest grains, therefore the smallest fragment size plays a larger role on determining the maximum fragment size. For the mean value of \change{$\alpha = 1.3$, $r_c = $  8 to 18 $\mu$m for $r_0 = $ 10 nm to 0.3 $\mu$m}{$\alpha = 1.1$, $r_c = $  21 to 30 $\mu$m for $r_0 = $ 10 nm to 0.3 $\mu$m} respectively.  \add{While we cannot empirically constrain $r_0$ in this setup, we investigate a lower limit much smaller than that of $\beta$-meteoroids. The derived mass-indices in this study, calculated from the slope of the measured voltage distributions, is insensitive to the choice of $r_0$. Hence, we include the discussion of $r_0$ for intuitive context and visualization of the consequences of different values of the mass index. }

\begin{figure*}
\plotone{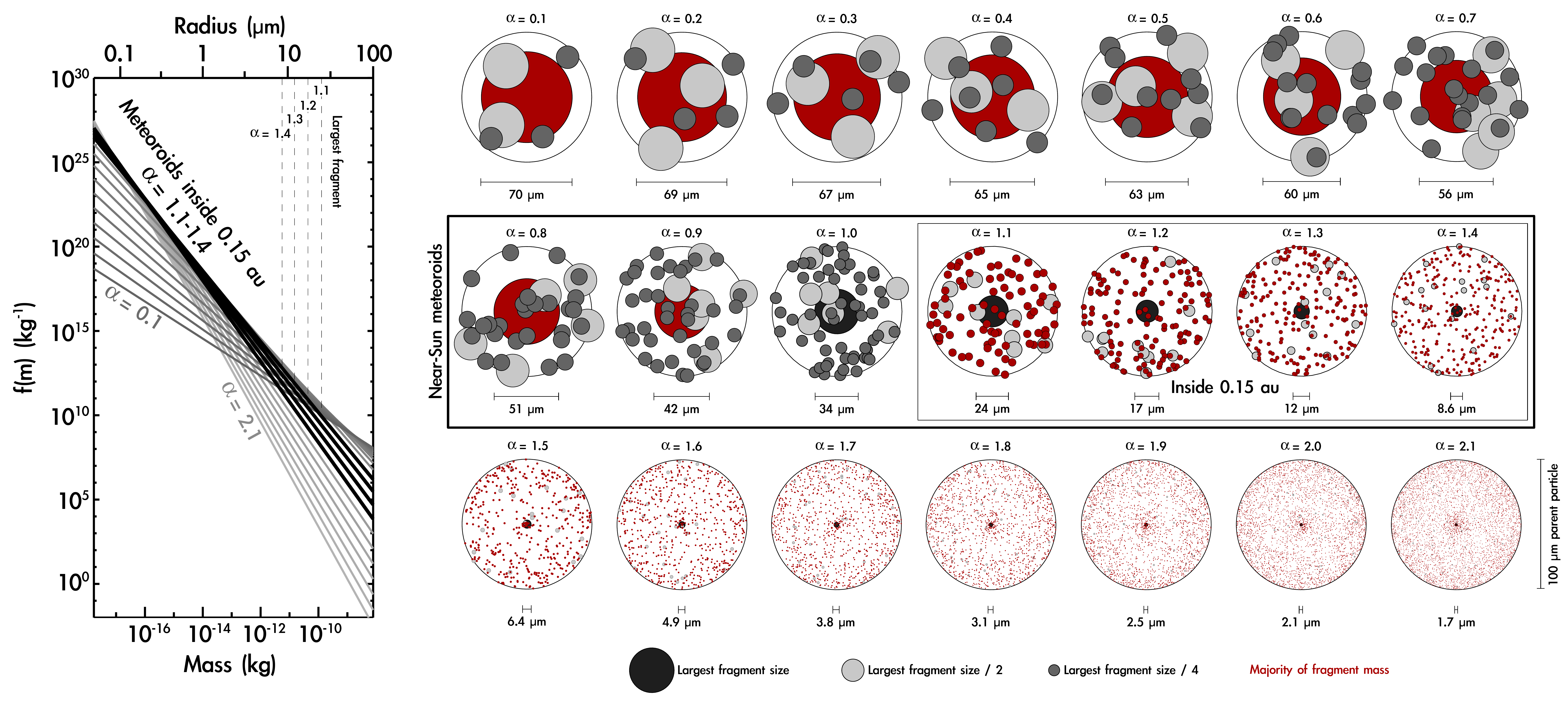}
\caption{Size distribution of collisional fragments. (left) Differential mass distributions, with $\alpha = 1.1 \pm 0.3$ highlighted in black and the largest fragment size marked with the vertical dashed lines. (right) Visual representation of fragments, with the single largest fragment in the center and the size of the largest size fragment denoted below each. The fragments that are 1/2 and 1/4 of the largest fragment size are shown in shades of grey, with the number of circles proportional their size distribution. Red indicates where the majority of fragment mass resides, where the largest grains carry the mass for $\alpha < 1$ and smallest grains carry the mass for $\alpha > 1$.
 \label{fig:power}}
\end{figure*}

To provide an intuitive understanding of how these mass indices affect the near-Sun dust distribution, we visualize the size distribution of collisionally produced grains with $r_0 = 50$ nm and $r_1 = 100 $ $\mu$m in Figure~$\ref{fig:power}$. This value of $r_1$ is chosen due to the fact that at 1 au, the majority of mass in the zodiacal cloud was found to be in $\sim$100 $\mu$m radius grains \citep{grun:85a}, hence this size grain is likely to be the parent of much of the zodiacal collisional products. Figure~$\ref{fig:power}$ shows the size distribution for $\alpha = 0.6 - 2.0$ with these assumptions. 

While the slopes of the different values of $\alpha$ are shown in the left portion of this figure, the schematic on the right visualizes how mass is distributed into collisional fragments. As previously discussed, most of the mass for $\alpha < 1.0$ is distributed into large grains and therefore collisional products in this regime have a few ``large'' collisional products that carry the majority of the fragmented material. However, in the regime for $\alpha > 1.0$, as found for the near-Sun environment, most of the mass gets partitioned into very small grains, while still allowing for a few larger grain fragments. \remove{For example, for the average value of $\alpha = 1.3$ found here, a 100 $\mu$m radius grain would fragment such that the largest byproduct is 12 $\mu$m.} With the exception of the few largest grains, the overwhelming majority of collisional mass is partitioned into the very lowest end near $m_0$, which for these size ranges would all become \bms to be ejected from the solar system. 

\begin{figure*}
\plotone{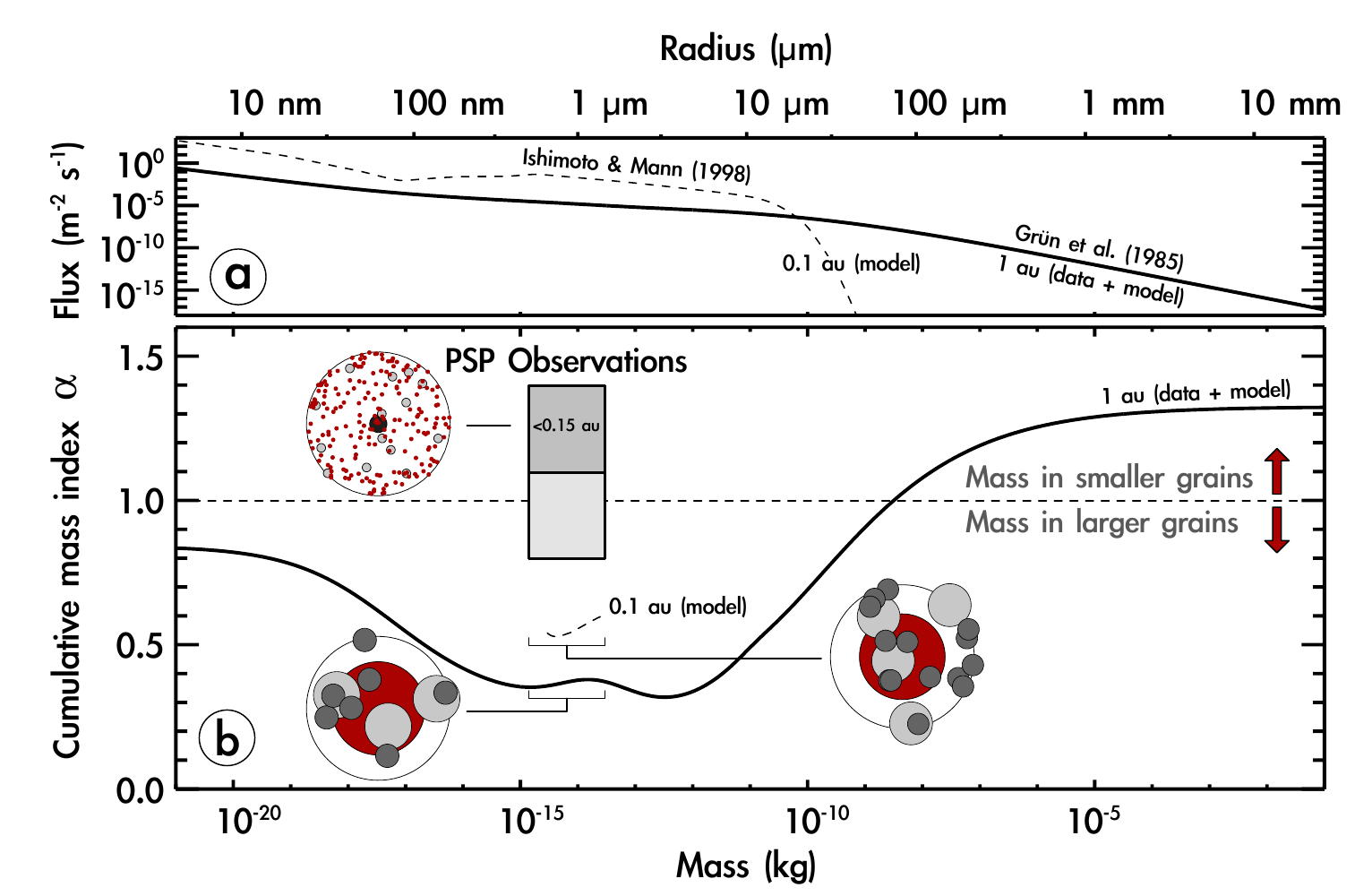}
\caption{Comparison with size distribution at 1 au \citep{grun:85a} derived from a large number of observations, as well as the modeled distribution at 0.1 au \citep{ishimoto:98a}. The grey boxes shows the values of $\alpha$ found here for grains with radii in the approximate range of $\sim$0.6-1.4$~\mu$m detectable by PSP. \label{fig:grun}}
\end{figure*}

To compare to expectations, Figure~\ref{fig:grun}a shows the cumulative flux of meteoroids $F(m)$ with masses $>m$ at 1 au \citep{grun:85a}. The cumulative mass index of this distribution can be calculated assuming for sufficiently small ranges in mass that the cumulative mass distribution $F(m) \propto m^{-\alpha}$, such that
\begin{equation}
\alpha = -\frac{m}{F(m)}\frac{\textrm{d}F(m)}{\textrm{d} m}
\end{equation}
and shown in Figure~\ref{fig:grun}b. Figure~\ref{fig:grun}b shows our derived mass indices for meteoroids from 0.1 to 0.25 au of \change{$\alpha = 1.3 \pm 0.2$}{$\alpha = 1.1 \pm 0.3$} with the approximate mass/size range detectable by PSP denoted by the width of the grey rectangle. The corresponding mass index at 1 au is $\alpha = 0.35-0.4$ \citep{grun:85a}, therefore the mass index very near the Sun is significantly larger than its corresponding 1 au value. We also show the modeled expectation of the zodiacal flux and mass index at 0.1 au \citep{ishimoto:98a} with the dashed lines in Figure~\ref{fig:grun}, where within the size range PSP can probe, the expected size indices at 0.1 au were expected to be $\alpha = 0.5-0.6$

Both the previous expected mass indices at 0.1 and 1.0 au were below $\alpha = 1$, therefore, they were both in the regime in which most of the collisional products were still large grains that would remain bound. This is in contrast to the values found here of  $\alpha = 1.3 \pm 0.2$ \add{inside 0.15 au}, where effectively all collisional products immediately become \bms to be expelled from the solar system. This discrepancy may be due to the particularly intense and high speed impact environment near the Sun.

 A similar technique was applied to analyze the distribution of dust impacts to SolO \citep{zaslavsky:21a}, which found values of $\alpha = 0.3-0.4$. However, the index derived from SolO impacts was likely for unbound \bms with radii of $\sim$100-200~nm, compared to the index calculated here for bound \ams of radii $\sim$0.6-1.4$~\mu$m. Additionally, the mass index derived from SolO observations was quoted as a lower bound that may be sensitive to the distribution of \bmm impact speeds which could have an appreciate spread in speed at SolO's location \citep{zaslavsky:21a}. Cross-comparing PSP and SolO impactor distributions for similar populations, particularly if \bms could be isolated in both, would provide an exciting opportunity to evaluate the evolution of the size distribution as a function of heliocentric distances. 
 
 The population of \bms was specifically excluded from the calculation of mass indices in this study to focus solely on \ams which should have a smaller spread in impact speed. If the power-law index found here holds down to \bmsn, future modeling efforts to understand how a size distribution of \bms produced very near the Sun could be compared to all existing and future measurements of \bms from PSP, SolO, STEREO, WIND, and/or any other spacecraft with the capability to detect \bms throughout the \change{innner}{inner} heliosphere \citep{juhasz:13a,obrien:18a,poppe:20a,poppe:22a}.

\section{Discussion and Conclusions} \label{sec:conclusions}
Leveraging PSP's eccentric orbit, we identify a region near the Sun from 0.1-0.25 au where PSP is effectively only observing bound \ams due to the very low relative impact speeds and corresponding fluxes from \bms in this region. From these observations, we derive the cumulative power-law mass index to be \change{$\alpha = 1.3 \pm 0.2$}{$\alpha = 1.1 \pm 0.3$} for grains with radii in the approximate range of $\sim$0.6-1.4$~\mu$m. These results quantitatively demonstrate the meteoroid size distribution near the Sun differs substantially from the distribution at 1 au. In contrast to expectations, we find most of the collisional fragments produced in the near-Sun environment inside 0.15 au are overwhelmingly partitioned into sub-micron \bmsn, which are effectively ``blown out'' of the solar system via the Sun's radiation pressure, or even smaller grains that could become electromagnetically coupled to the solar magnetic field \citep{czechowski:12a}. The mass index also increases for decreasing heliocentric distance, suggesting even smaller fragments are produced closer to the Sun. The discrepancy between the lower expected mass index and higher value found here could be due to the intense fragmentation process occurring in the very near-Sun environment, where impact speeds between grains are likely the highest in the solar system and may provide significantly different fragmentation products compared to lower-speed impacts farther from the Sun. The canonical mass flux at 1 au from \citet{grun:85a} is often scaled by $r^{-0.5}$ to estimate the dust flux at different heliocentric distances; these findings demonstrate that the 1 au fluxes cannot be simply scaled to estimate the near-Sun dust distribution. 

For this analysis, we assume collisions are driving the size distribution near the Sun. \add{Furthermore, to interpret the mass indices derived here we assume collisional fragmentation produces only dust grains as fragmentation products and consider the production of impact vapor/plasma negligible compared to the mass of the fragments. While modeling efforts have been investigated on grain destruction and the role of vaporization} \citep[e.g.][]{borkowski:95a}, \add{we elected not to include more complex origin mechanisms in order to interpret the derived mass-indices with minimal assumptions, which future analyses could explore.

In addition to collisions,} sublimation is certainly also operating on grains that transit very near the Sun, where grains are expected to strongly sublimate within $\sim$10 $\rs$ \citep{mann:04a}. However, the location of peak production of \bms from a collision model was found to be consistent with the derived \bmm fluxes to PSP, such that collisions were inferred to be the driving phenomenon for dust grain populations observed by PSP \citep{szalay:21a}. Additionally, a comprehensive dynamical meteoroid model which \change{include}{includes} fragmentation and the production of \bms was found to well-match the PSP observations of \bmm fluxes in the inner heliosphere \citep{pokorny:24a}.

\add{Additionally, we assume that transport effects have not significantly modified the collisionally produced size distribution from their sources such that the collisional fragmentation index $\eta = \alpha$. We anticipate most of the collisions occur below 0.1 au} \citep{szalay:21a}, \add{such that the small $\alpha$-meteoroids we measure outside this distance are likely to have been produced at similar radial distances and may not have been significantly evolved from their original source mass distributions. Future efforts could investigate the effects of transport, specifically how the mass-dependent Poynting-Robertson drag would further modify collisional mass distributions.}

Rotational disruption due to radiative torque from sunlight on irregular grains has also been proposed to play an important role for dust evolution within our solar system and in astrophysical systems \citep{misconi:93a,silsbee:16a,hoang:19a}. While grains released from comets have been found to be imparted with rotation due to gas drag \citep{capek:14a}, the extent to which grains experience radiative torques has not been observationally constrained in the zodiacal cloud. While we cannot assess the extent to which rotational disruption plays a role in the zodiacal dust evolution, the size distribution found here provides constraints for any model that would predict the mass partitioning of rotationally disrupted grains and their relative contribution to the overall zodiacal dust population.

Particles originating from short-period comets have been shown to be resilient to collisions with the zodiacal cloud and have higher fragmentation strength than commonly assumed \citep{pokorny:24a}. This translates to longer collisional lifetimes, the ability of larger particles to migrate closer to the Sun, and a smaller fraction of interplanetary dust particles lost to collisions. The higher particle collisional strength might explain the discrepancy in the value of $\alpha$ between our work and existing models \citep{grun:85a, ishimoto:98a}, as shown in Figure~\ref{fig:grun}. More particles are able to migrate into the very inner heliosphere to then be detected by PSP without being collisionally fragmented into nano-dust particles that are not detectable by PSP instruments. Our results might provide a crucial missing piece of evidence for determining the crushing laws in the zodiacal cloud and for models of collisional evolution of interplanetary dust clouds \citep{pokorny:24a, poppe:16c,poppe:19a}.

We also note that a fundamental dust grain unit size on the order of 10's of $\mu$m was found from observations of cometary dust from 67/C-G, interpreted to be the remnants of the early accretion processes \citep{hornung:23a}.
If there are grains that represent more `fundamental units', they could be more resilient to fragmentation than their aggregates. The \citet{grun:85a} curve in Figure \ref{fig:grun} shows a steep decline in the value of $\alpha$ once the particle radius drops below $\sim$$100~\mu$m, perhaps suggestive that grains of this size are more resilient to fragmentation. While \citet{hornung:23a} report their findings using the in-situ experiment on 69P/C-G, the \citet{grun:85a} model is based on spacecraft observations from HEOS2, Pioneers 8 and 9, and from lunar microcraters, all at 1 au. Interplanetary dust particles experience complex space weathering during their lifetime which influences their chemical composition as well as their size-frequency distribution. While the discrepancy is apparent, how grain fragments relate to the original makeup of their parent grains cannot be directly determined from this analysis, yet provides an interesting topic for further investigation.


These results can also be applicable to exozodiacal disks. The grain size distribution, as well as the local stellar conditions at exozodiacal systems, controls the total accretion rate of dust onto exoplanets \citep{arras:22a}. The relative partitioning into \ams and \bms in exozodiacal disks plays an important role in their collisional evolution and  dictates how remote measurements from these disks are interpreted \citep{krivov:00a}. For example, it was found that \bmm fluxes within the debris disk of $\beta$ Pictoris could be so large they significantly contribute to the fragmentation of \amsn, hence they may modify the size distribution of their parent \ams and govern the size dominating in overall cross-section that would be remotely detectable \citep{krivov:00a}. \bms were also found to play an important role in the luminosities of debris disk halos, where an updated size-frequency distribution could aid in the interpretation of optical and near-IR observations of these halos \citep{thebault:23a}.

Summarizing our results,
\begin{itemize}
    \item The cumulative mass index for grains with radii of $\sim$0.6-1.4 $\mu$m (masses of $3\times10^{-15}$ to $3\times10^{-14}$ kg) is found to be \change{$\alpha = 1.3 \pm 0.2$}{$\alpha = 1.1 \pm 0.3$} from 0.1-0.25 \rst 
    \item The cumulative mass index increases \remove{monotonically} toward the Sun, with even smaller collisional fragments generated closer to the Sun. 
    \item The derived size distribution is steeper than previously estimated, and in contrast to expectations we find most of the dust mass resides in the smallest fragments \add{inside 0.15 au} and not in large grains. 
    \item These results place new constraints how the solar system's zodiacal cloud and by extension astrophysical debris disks are partitioned in mass.
\end{itemize}

In conclusion, PSP observations provide a unique platform to observe the inner-most regions of our zodiacal dust disk. While PSP does not have a dedicated dust instrument, impacts detected by its electric field instrument can be used to quantitatively constrain the multiple populations in the near-Sun dust environment. These observations are taken in the nearest region to the Sun ever explored in-situ by spacecraft and provide a critical window into the complex evolution of our zodiacal cloud. 

\textbf{Acknowledgements.} We thank the many Parker Solar Probe team members that enabled these observations. We thank B. Draine, M. Horanyi, B. Hensley, \& K. Silsbee for helpful discussions on grain fragmentation, D. McComas \& G. Livadiotis for helpful discussions related to the power-law analyses, and D. P. Morgan \& S. P. Childress for graphical guidance. We also thank the two anonymous reviewers for their helpful comments during review. We acknowledge NASA Parker Solar Probe Guest Investigator grant 80NSSC21K1764. P.P. was additionally supported by NASA Solar System Workings award number 80NSSC21K0153, a cooperative agreement 80GSFC21M0002, and NASA’s Planetary Science Division Research Program, through ISFM work packages EIMM and Planetary Geodesy at NASA Goddard Space Flight Center. We used the ``managua'' \& ``bukavu'' colormaps \citep{crameri:20a} for Figures~\ref{fig:index} \& \ref{fig:alpha}.

%






\appendix
\section{Constraining the Uncertainty in Estimating the Mass Index}

When a high-speed dust grain with mass $m_\mathrm{imp}$ and speed $v_\mathrm{imp}$ impacts a surface (PSP) it produces an impact charge $Q_\mathrm{imp} \propto m_\mathrm{imp}^a v_\mathrm{imp}^b$, where the exponents $a$, $b$ and proportionality constant are determined empirically based on the surface properties \citep{auer:01a}. Typical spacecraft surface materials have a mass exponent of 1 \citep{collette:14a}, and we adopt this value here. \remove{hence, we assume the impact charge produced from dust impacts to PSP is directly proportional to their mass}. \add{For the speed dependence, we assume the impact speed has a functional form of $v_\mathrm{imp} = \bar{v}_\mathrm{imp} m^{c}_\mathrm{imp}$, such that there is a separable component of the impact speed that depends on mass and $\bar{v}_\mathrm{imp}$ has no mass dependence. Combining the empirically determined mass exponent with a value of 1.0 with the additional mass dependence from the impact speed, the mass dependent exponent of the impact charge is $a = 1+\delta a$, where $\delta a = bc$}

\remove{therefore for grains impacting PSP at the same speed, the impact amplitude $\Delta V$ is proportional to $m_\mathrm{imp}$}. \remove{As previously discussed, the impact charge produced by impacts has a dependence on the impact speed such that $Q_\mathrm{imp} \propto m_\mathrm{imp}^a v_\mathrm{imp}^b$.} \change{For the most likely target materials on the spacecraft to be hit by dust and produce impact signals, empirically determined exponents for Aluminum and MLI (multilayer thermal insulation) are $b=3.48$ and $b=4.7$ respectively}{The velocity index $b$ has been empirically determined for a variety of spacecraft materials. MLI (multilayer thermal insulation) covers much of the spacecraft and is the most likely target material to be hit by dust and produce impact signals for bound grains post perihelion that would impact the spacecraft body instead of the heat shield} \citep{szalay:20a}. We therefore assuming an exponent of $b=4.7$ for MLI \citep{collette:14b}. 

\add{With such a relation between $Q$ and $m$, the cumulative charge index is given by $\alpha_Q = \alpha/a  =\alpha/(1+\delta a)$. Therefore, the cumulative mass index is related to the cumulative impact charge index by}
\begin{equation}
\alpha = \alpha_Q(1+\delta a)
\end{equation} 
\add{To calculate the mass-dependent component of the impact speed, $\delta a$, we investigate the dependence of the dust impact speed to PSP. For motion in a two-dimensional orbital plane, a body in a bound orbit about the Sun has an orbital velocity such that}
%

\begin{equation*}
v = \sqrt{\mu(1-\beta) \left(\frac{2}{r}-\frac{1}{a} \right)} 
\end{equation*}

\begin{equation*}
v_\varphi = \frac{\sqrt{\mu(1-\beta)a(1-e^2)}}{r}
\end{equation*}

\begin{equation*}
v_r = \pm \sqrt{  \mu(1-\beta) \left( \frac{2}{r}-\frac{1}{a}-\frac{a(1-e^2)}{r^2}  \right)   }
\end{equation*}

\add{where $\mu = GM_\odot = 1.327 \times 10^{20}$ m$^3$ s$^{-2}$ is the standard solar gravitational parameter, $\beta$ is the mass-dependent ratio of the radiation pressure and gravitational forces from the Sun} \citep{zook:75a,burns:79a,wilck:96a}, \add{$r$ is the heliocentric distance, $e$ is the eccentricity, and $a$ is the semi-major axis. These equations also apply to PSP with $\beta=0$.}

\begin{figure*}
\plotone{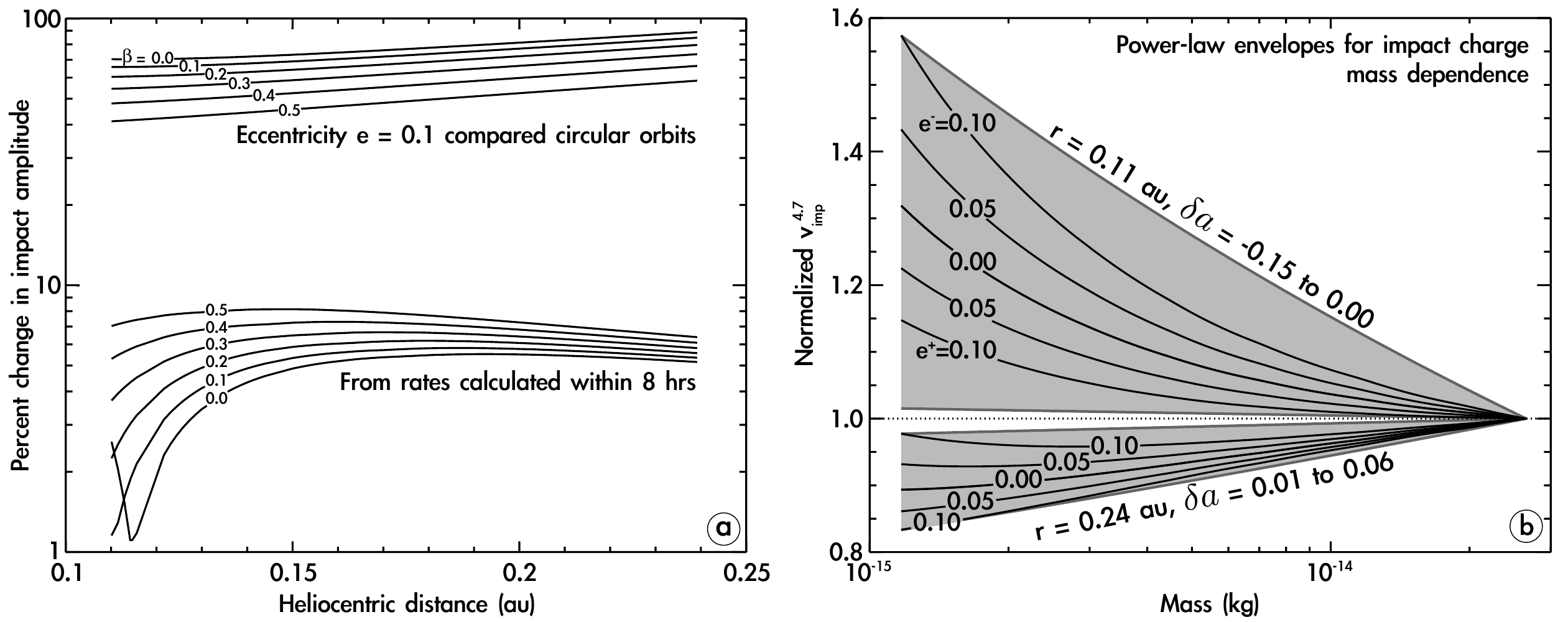}
\caption{(a) Percent change in the impact amplitude due to the finite cadence of 8 hr impact rate intervals and due to the possible eccentricity of the grains. For each effect, six curves are shown corresponding to $\beta = 0-0.5$.  (b) Power-law envelopes to the family of normalized mass-dependent impact charge curves for heliocentric distances of 0.11 au and 0.24 au. \label{fig:unc}}
\end{figure*}

\add{To reduce complexity in the following calculation, we assume each impacting grain is on a Keplerian orbit with a semi-major axis equal to the heliocentric distance of PSP at any location, such that $a_d = r$, and that grains are in the same orbital plane as PSP. With these assumptions, the dust impact speed $v_\textrm{imp} = |\vec{v}_\mathrm{d}-\vec{v}_\mathrm{PSP}|$ to PSP is}
\begin{widetext}
\begin{equation}
v_\textrm{imp} = \sqrt{\mu \left[ 
\frac{2}{r} 
- \frac{1}{a} 
+ \frac{1-\beta}{r} 
- 2\sqrt{\frac{(1-\beta)a(1-e^2)(1-e_d^2)}{r^3}} 
\pm 2e_d\sqrt{\frac{(1-\beta)}{r}  \left(  \frac{2}{r}-\frac{1}{a}-\frac{a(1-e^2)}{r^2}  \right)  }
 \right]}
 \label{eq:vimp}
\end{equation}
\end{widetext}
\add{where the mass dependence is from $\beta(m)$, $e_d$ is the eccentricity of the dust, $a = 0.41$ au \& $e=0.85$ are the semi-major axis and eccentricity of PSP during orbits 10-16. The $\pm$ has a positive value when PSP and the dust grain have the opposite sign for their radial velocity components (e.g. post-perihelion outbound PSP and inbound dust grains), and negative when the radial velocity components have the same sign. This impact speed relation will allow us to constrain the uncertainties of our estimates for $\alpha$ discussed at the end of this section.}


 \add{With the analytic expression for the impact speed to PSP given in} Eq.~\ref{eq:vimp} \add{we can estimate the uncertainty in determining the mass index due to three effects that all vary as a function of heliocentric distance. First, all impact rates are calculated in 8-hour intervals, which introduces a small uncertainty on the impact speeds.} Figure~\ref{fig:unc}a \add{shows the percent change in impact charge amplitude calculated via $(1+\Delta v/v)^{4.7}$ due to this effect, where $v$ is the impact speed to PSP from a grain on a circular, Keplerian orbit in the center of an 8-hr interval for each distance and $\Delta v$ is the maximum expected difference between $v$ and the impact speeds within the entire 8-hr interval. For the region investigated, this introduces an impact amplitude uncertainty of $<$10\% and therefore we neglect this effect.}

\add{The next effect we investigate is possible uncertainty in impact speed due to grains having a non-zero eccentricity. Zodiacal grains very near the Sun are expected to have low eccentricities due to circularization from Poynting-Robertson drag and we consider the case for grains that have eccentricities of up to 0.1} \citep{pokorny:24a}. Figure~\ref{fig:unc}a \add{shows the percent change in impact charge amplitude due to such non-circular impactors and we find this introduces an uncertainty in the impact charge amplitude of up to a factor of 2 which translates to a factor of $\sqrt[3]{2} \approx 1.3$ in radius. As grains on circular orbits would have the slowest impact speed compared to eccentric grains with the same other orbital parameters, this uncertainty means we may be overestimating the impactor grain size by a factor of $\sim$1.3.} 

\add{Both of these uncertainties do not introduce any mass-dependent bias and therefore would not affect the determination of the mass index, they would only affect the estimate of the relevant size range to which the mass index is valid. The third source of uncertainty we consider is due to the impact speed having a mass dependence, which does introduce a mass-dependent bias and propagates to our estimates of $\alpha = \alpha_Q(1+\delta a)$.}

\add{To quantify $\delta a$, we calculate the impact speed to PSP given in} Eq.~\ref{eq:vimp} \add{as a function of mass, where the mass dependence is solely from $\beta(m)$. We use the $\beta(m)$ relation corresponding to ``old cometary'' grains according to} \citet{wilck:96a}. \add{For the size range considered, $\beta$ has values near 0.5 on the lowest mass end and 0.2 on the upper mass end.} Figure~\ref{fig:unc}b \add{shows two sets of normalized $v_\textrm{imp}^{4.7}$ curves corresponding to the heliocentric distance boundaries of 0.11 au and 0.24 au considered here. Multiple curves for each family are shown for different values of eccentricity between 0 and 0.1 and using the $+$ or $-$ sign in} Eq.~\ref{eq:vimp}, \add{with the curves corresponding to a minus sign as the top two of each family and those corresponding to a plus sign as the bottom two. We then determine the slope of power-laws that provide an envelope to these family of curves for each heliocentric distance. In the examples given in }Figure~\ref{fig:unc}b, \add{we find power-law functional forms with $\delta a =$ -0.15 to 0 provide an envelope for the family of curves corresponding to $r=0.11$ au and $\delta a =$ 0.01 to 0.06 for $r = 0.24$ au.} \add{We calculate these envelopes over the entire range of heliocentric distances considered in this study to determine the range of $\delta a$ minimum and maximum values.}


\bibliographystyle{aasjournal}



\end{document}